\documentclass[prb,showpacs,print,twocolumn]{revtex4}
\usepackage{pifont}
\usepackage{amsmath}
\usepackage{graphicx}
\usepackage{dcolumn}
\usepackage{bm}
\usepackage{txfonts}

\begin{document}

\title{Landau levels and magneto-transport property of monolayer phosphorene}

\author{X. Y. Zhou,$^{1,2}$ R. Zhang,$^1$ J. P. Sun,$^1$ Y. L. Zou,$^1$ D. Zhang,$^1$ W. K. Lou,$^1$ F. Cheng,$^4$ G. H. Zhou,$^2$ F. Zhai,$^3$ Kai Chang$^{1}$\footnote{To whom correspondence should be addressed: kchang@semi.ac.cn}}

\affiliation{$^1$SKLSM, Institute of Semiconductors, Chinese Academy
of Sciences, P.O. Box 912, Beijing 100083, China.}

\affiliation{$^2$Department of Physics and Key Laboratory for
Low-Dimensional Structures and Quantum Manipulation (Ministry of
Education), Hunan Normal University, Changsha 410081, China.}

\affiliation{$^3$Department of Physics, Zhejiang Normal University,
Jinhua 321004, China.}

\affiliation{$^4$Department of Physics and Electronic Science,
Changsha University of Science and Technology, Changsha 410004,
China.}

\begin{abstract}
We investigate theoretically the Landau levels (LLs) and
magneto-transport properties  of phosphorene under a perpendicular
magnetic field within the framework of the effective
\textbf{\emph{k$\cdot$p}} Hamiltonian and tight-binding (TB) model.
At low field regime, we find that the LLs linearly depend both on
the LL index $n$ and magnetic field $B$, which is similar with that
of conventional semiconductor two-dimensional electron gas. The
Landau splittings of conduction and valence band are different and
the wavefunctions corresponding to the LLs are strongly anisotropic
due to the different anisotropic effective masses.
An analytical expression for the LLs in low energy
regime is obtained via solving the decoupled Hamiltonian, which
agrees well with the numerical calculations. At high magnetic
regime, a self-similar Hofstadter butterfly (HB) spectrum is
obtained by using the TB model. The HB spectrum is consistent with
the Landau level fan calculated from the effective
\textbf{\emph{k$\cdot$p}} theory in a wide regime of magnetic
fields. We find the LLs of phosphorene nanoribbon depend
strongly on the ribbon orientation due to the anisotropic hopping parameters.
The Hall and the longitudinal conductances (resistances)
clearly reveal the structure of LLs.
\end{abstract}
 \vspace{0.2cm}
\maketitle

\section{Introduction}
The group V element phosphorus can exist in several allotropes and
black phosphorus (BP) is the most stable phase under normal
conditions.\cite{Nishii} Recently, layered BP has attracted
intensive attention because of its unique electronic properties and
potential applications in
nanoelectronics.\cite{Rodin,YBZhang,Han,Ye,Churchill,Reich,Andres}
In the bulk form, BP is a van der Waals-bonded layered material where
each layer forms a puckered surface due to $sp^3$
hybridization.\cite{Rodin,YBZhang} BP possesses a direct band gap
0.3 eV located at Z point.\cite{YBZhang,Han} This direct gap moves
to $\Gamma$ point and increases to 1.5-2 eV when the thickness
decreases from bulk to few layers and eventually monolayer via
mechanical exfoliation.\cite{YBZhang,Ye,Lu} Hence, BP is an
appealing candidate for tunable photodetection from the visible to
the infrared part of the spectrum.\cite{Buscema} Further, the
field-effect-transistor (FET) based on few layer BP is found to have
an on/off ratio of 10$^5$ and a carrier mobility at room temperature
as high as 10$^3$ cm$^2$/V$\cdot$s,\cite{YBZhang,Ye,Tayari} which make BP a
favorable material for next generation electronics.

The low energy physics of monolayer BP (phosphorene) around $\Gamma$ point can be
well described by an anisotropic two band \textbf{\emph{k$\cdot$p}}
model,\cite{Rodin} which agrees well with a tight binding (TB)
model.\cite{Rudenko} To date, various interesting properties for
phosphorene have been predicted theoretically and verified experimentally, including those related to strain
induced gap modification,\cite{Rodin} tunable optical
properties,\cite{Tony2} layer controlled anisotropic
excitons,\cite{Tran} quantum oscillations in few layers BP\cite{YBZhange,XLChen,Tayari} etc.
However, the Landau levels (LLs) and magneto-transport (MT) properties of this unique anisotropic system
remain unexplored.

In this work, we study the LL spectra and MT properties of
phosphorene under a perpendicular magnetic field. By using an
effective \textbf{\emph{k$\cdot$p}} Hamiltonian, we find that the
LLs linearly depend both on energy index $n$ and magnetic field $B$
at low-field regime, which means the LLs in phosphorene are similar
with that in conventional semiconductor two dimensional gases
(2DEGs). Interestingly, owing to the anisotropic energy dispersions,
i.e., the effective masses, the Landau splittings of conduction and
valence band are different for a fixed magnetic field, and the
wavefunctions corresponding to the LLs show strong anisotropic
behavior. We obtain an analytical expression for the LLs in low
energy regime via solving a decoupled Hamiltonian, which agrees well with
the numerical data in low energy regime. At high-field regime,
magneto-level spectrum, i.e., the Hofstader butterfly (HB) spectrum,
is obtained by using a tight binding (TB) model. We find that the
results obtained by the effective \textbf{\emph{k$\cdot$p}}
Hamiltonian and TB model agree with each other in weak magnetic
field cases. Further, we find the LLs of phosphorene nanoribbon
depend strongly on the ribbon orientation due to the anisotropic hopping parameters.
In order to detect those interesting magneto energy
spectra, we calculate MT properties of phosphorene within the
framework of the linear response theory. By using Kubo formula, we
find the Hall and the longitudinal conductances (resistances) clearly reveal the
structure of LLs.

The paper is organized as follows. In Sec. II. A, we present the
calculation method and obtain the effective Hamiltonian and LL
spectra. In Sec. II. B, we calculate the Hall and longitudinal
conductance by using Kubo formula. In Sec. III, we present the
numerical results and discussions. Finally, we summarize our results
in Sec. IV.

\section{Model and Hamiltonian}

\subsection{Landau levels in monolayer phosphorene}
In this subsection, we present the effective
\textbf{\emph{k$\cdot$p}} and TB Hamiltonians and the calculation
method about the LLs at low and high magnetic fields. In the top
view of phosphorene, as shown in Figure 1(a), $\bm{a_1}$=3.32 $\AA$
and $\bm{a_2}$=4.48 $\AA$ are the primitive vectors, $a$=2.22 $\AA$
and $\theta$=96.76$^\text{o}$ are the in-plane bond length and bond
angle,\cite{Rudenko} respectively. The unit cell of phosphorene
contains four atoms (see the solid rectangle) with two phosphorus
atoms in the lower layer and the other two atoms in the upper layer.
Very recently, a tight binding (TB) model of phosphorene has been
proposed and is given by\cite{Rudenko}
\begin{eqnarray}
H=\sum_{<i,j>}t_{ij}c_j^{\dag}c_i,
\end{eqnarray}
where the summation runs over all the lattice sites of phosphorene,
$c_j^{\dag}(c_i)$ is the creation (annihilation) operator of electron
on the site $j(i)$, and $t_{ij}$ are the hopping parameters. It has been
shown that five hopping links [see Fig. 1(a)] are enough to describe
the electronic band structure of phosphorene.\cite{Rudenko} The
related hopping parameters are: $t_1$=$-$1.22 eV, $t_2$=3.665 eV,
$t_3$=$-$0.205 eV, $t_4$=$-$0.105 eV, and $t_5$=$-$0.055 eV.

Generally, the energy dispersion of phosphorene should be described
by a four band model\cite{Rudenko,Ezawa} in the TB framework.
However, it can be also expressed by a two-band model due to the
$C_{2h}$ point group invariance.\cite{Ezawa} In the two-band model,
the unit cell contains two phosphorus atoms [see the dashed
rectangle in Fig. 1(a)], where one is in the upper layer and the
other in the lower layer. Expanding the TB model around $\Gamma$
point with a coordinate
rotation\cite{Ezawa}($\tau_x$$\rightarrow$$\tau_z$,
$\tau_y$$\rightarrow$$\tau_x$), one obtains the low energy
\textbf{\emph{k$\cdot$p}} model for phosphorene, which reads
\begin{eqnarray}
H=\left({\begin{array}{*{20}c}
h_c &h_{cv} \\
h_{cv}^* &h_v
\end{array}}\right)=\left({\begin{array}{*{20}c}
E_c+\alpha k_x^2+\beta k_y^2 &\gamma k_x \\
\gamma k_x &E_v-\lambda k_x^2 -\eta k_y^2
\end{array}}\right),
\end{eqnarray}
where $E_c$=0.34 eV ($E_v$=$-$1.18 eV) is the conduction (valence) band
edge, $\gamma$=$-$5.2305 eV$\cdot\AA$ describes the interband
coupling between the conduction and valence band, parameters
$\alpha$, $\beta$, $\lambda$, $\eta$ are related to the effective
masses with $\alpha$=$\hbar^2/2m_{cx}$, $\beta$=$\hbar^2/2m_{cy}$,
$\lambda$=$\hbar^2/2m_{vx}$, $\eta$=$\hbar^2/2m_{vy}$. Here
$m_{cx}$=0.793$m_e$, $m_{cy}$=0.848$m_e$, $m_{vx}$=1.363$m_e$,
$m_{vy}$=1.142$m_e$, and $m_e$ is the free electron mass. The eigenvalue
of this Hamiltonian is
\begin{eqnarray}
E_{\pm}=\frac{1}{2}[h_c+h_v\pm\sqrt{(h_c-h_v)^2+4\gamma^2k_x^2}],
\end{eqnarray}
where $+/-$ is for conduction/valence band, respectively. Similar with
other low energy \textbf{\emph{k$\cdot$p}} models,\cite{Rodin,Tony2}
the dispersion described by Eq. (3) is strongly anisotropic. The
energy gap $E_g$ is $E_c-E_v$=1.52 eV, which is consistent with the first principle calculations (2 to 2.2 eV)\cite{Tran} and also in line with the recently measured optical gap\cite{Ye} 1.45 eV (the quasiparticle band gap minus the exciton binding energy). Figure 1 presents the dispersion of TB (the black solid line) and the
\textbf{\emph{k$\cdot$p}} models (the red dashed line), from which
we find they agree well with each other in a quite wide energy
regime. It seems the energy dispersion is linearly along
$\Gamma$$-$$X$ direction [see Fig. 1(c)]. However, it is actually parabolic.
In the long wave limit, we have $(h_c-h_v)^2\approx E_g^2\gg4\gamma^2k_x^2$.
Hence, we can expand Eq. (3) and obtain the energy dispersion of conduction and
valence band, which reads
\begin{eqnarray}
E_+\approx h_c+\frac{\gamma^2k_x^2}{E_g}, E_-\approx
h_v-\frac{\gamma^2k_x^2}{E_g},
\end{eqnarray}
From Eq. (4), one can easily find that the dispersion near $\Gamma$
point is quadratic. Owing to the interband coupling, the effective
masses along $\Gamma$$-$$X$ direction are modified as
$m_{cx}'$=$\hbar^2/2(\alpha+\gamma^2/E_g)$=0.167$m_e$,
$m_{vx}'$=$\hbar^2/2(\lambda+\gamma^2/E_g)$=0.184$m_e$. However, the
effective masses along $\Gamma$$-$$Y$ remain unchanged with
$m_{cy}$=0.848$m_e$, $m_{vy}$=1.142$m_e$.
\begin{figure}
\includegraphics[width=0.5\textwidth]{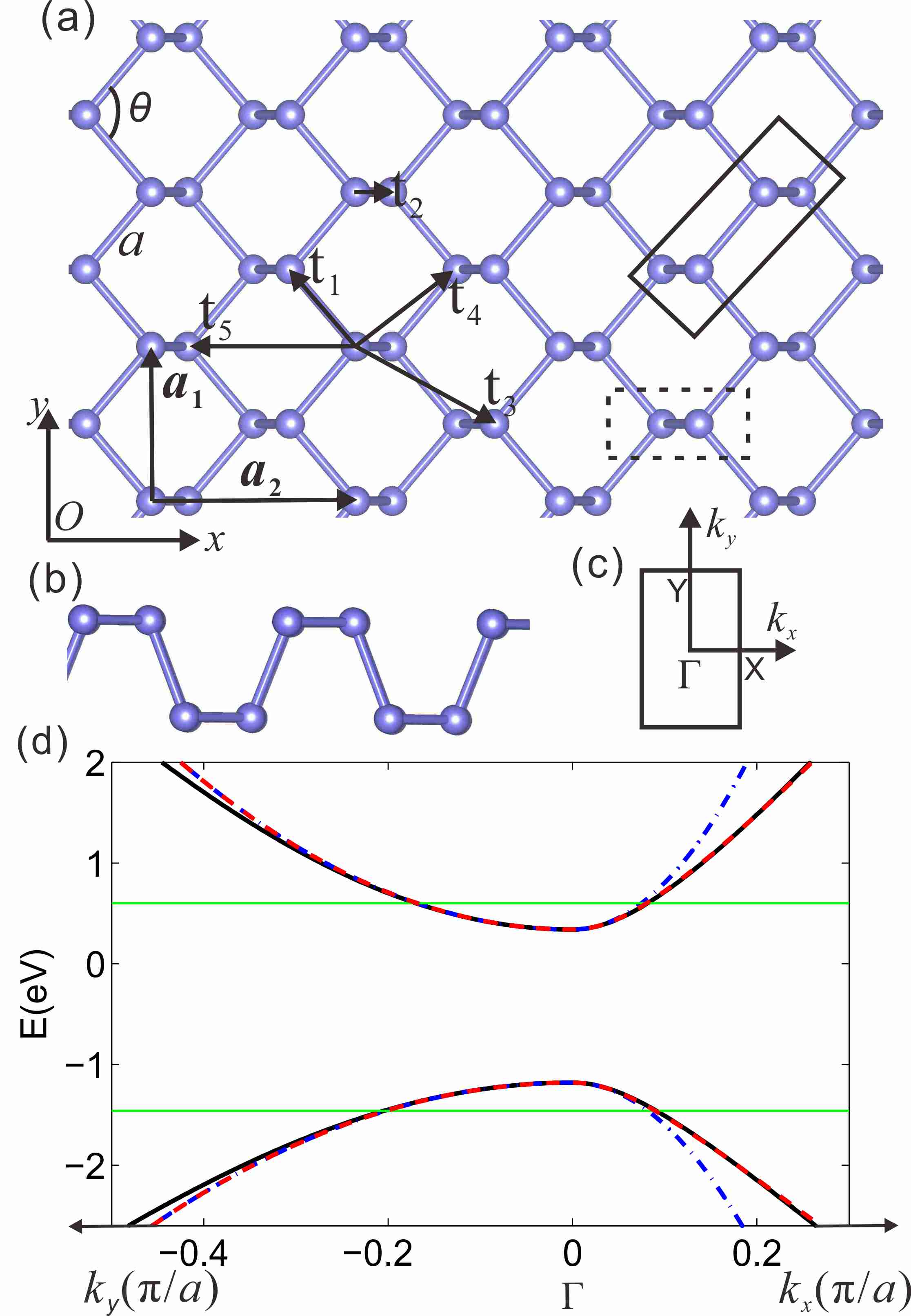}
\caption{(Color online) (a) The top view of phosphorene, $a$=2.22
$\AA$ ($\theta$=96.76$^\text{o}$) is the in plane bond length
(angel), $\bm{a_1}$ (3.32 $\AA$) and $\bm{a_2}$ (4.48 $\AA$) the
primitive vectors, $t_{i}(i=1,2,3,4,5)$ the five hopping links for
TB model. (b) The side view of phosphorene. (c) The first Brillouin
zone of phosphorene. (d) The energy dispersions of phosphorene, the
black solid and red dashed lines, represent the results obtained
from the TB and low energy \textbf{\emph{k$\cdot$p}} models,
respectively. The blue dash-dotted lines represent the results from
the decoupled Hamiltonian (13) and the green solid line illustrates
the energy regime where three Hamiltonians agree well with each
other.}
\end{figure}

When a perpendicular magnetic filed $\bm{B}$=$(0,0,B)$ is applied, we define the creation and annihilation operators as
\begin{eqnarray}
\hat{a}=\sqrt{\frac{m_{cy}\omega_c}{2\hbar}}(y-y_0+i\frac{p_y}{m_{cy}\omega_c}),\nonumber\\
\hat{a^{\dag}}=\sqrt{\frac{m_{cy}\omega_c}{2\hbar}}(y-y_0-i\frac{p_y}{m_{cy}\omega_c}),
\end{eqnarray}
where $\omega_c$=$eB/(m_{cx}m_{cy})^{\frac{1}{2}}$ is the frequency,
$y_0$=$l_B^2k_x$ is the cyclotron center, and $l_B$=$\sqrt{\hbar/eB}$
is the magnetic length. One finds Hamiltonian (2) turns to
\begin{eqnarray}
H=\left({\begin{array}{*{20}c} h_c &0\\ 0
&h_v\end{array}}\right)+h_R+h_D,
\end{eqnarray}
with
\begin{eqnarray}
h_R&=&\hbar\omega_{\gamma}\left({\begin{array}{*{20}c} 0 &\hat{a}\\
\hat{a}^{\dag} &0\end{array}}\right),h_D=\hbar\omega_{\gamma}\left({\begin{array}{*{20}c} 0 &\hat{a}^{\dag}\\
\hat{a} &0\end{array}}\right),\nonumber\\
h_c&=&E_c+(\hat{a}^{\dag}\hat{a}+1/2)\hbar\omega_c,\nonumber\\
h_v&=&E_v-(\hat{a}^{\dag}\hat{a}+1/2)\hbar\omega_v-(\hat{a}^2+\hat{a}^{\dag
2})\hbar\omega',
\end{eqnarray}
where $\omega_{\gamma}$=$\gamma/\sqrt{2}\hbar l_B\alpha_{yx}$,
$\omega_v$=($r_x$+$r_y$)$\omega_c$,
$\omega'$=($r_x$-$r_y$)$\omega_c/2$, with
$\alpha_{yx}$=$(m_{cy}/m_{cx})^{\frac{1}{4}}$,
$r_x$=$m_{cx}/2m_{vx}$ and $r_y$=$m_{cy}/2m_{vy}$. Interestingly,
the second (third) term in Eq. (6) looks like the Rasshba
(Dressehaus) spin-orbit interaction in conventional semiconductor
2DEG.\cite{Yang} In order to understand how the non-diagonal element $h_R$ and $h_D$
couple the Landau levels (LLs) in conduction and valence band, we first simplify the
Hamiltonian by ignoring the third term in $h_v$ [see Eq. (7)] since it is a
second-order perturbation. It will be included in numerical calculation. In this approximation,
we see that the term $h_R$ couples the LL $\phi_{n}|c\rangle$ with
$\phi_{n+1}|v\rangle$, while $h_D$ couples $\phi_{n+1}|c\rangle$
with $\phi_{n}|v\rangle$, where $|c\rangle=\binom10$,
$|v\rangle=\binom01$, $\{\phi_{n}\}$ are wave functions of the harmonic oscillator corresponding to $h_c$.

Taking Landau gauge $\bm{A}$=$(-By,0,0)$, when only the term $h_R$
exits, we obtain $E_{n,\pm}^R$=$(E_{nc}+E_{nv}\pm\Omega_n^R)/2$,
$\psi_{nk_x\pm}^R$=$e^{ik_xx}/\sqrt{L_x}\varphi_{nk_x\pm}^R$
($n$=1,2,...), where
\begin{eqnarray}
\varphi_{nk_x+}^R=\sin\frac{\vartheta_n}{2}\phi_{n-1}|c\rangle+\cos\frac{\vartheta_n}{2}\phi_{n}|v\rangle,\nonumber\\
\varphi_{nk_x-}^R=\cos\frac{\vartheta_n}{2}\phi_{n-1}|c\rangle-\sin\frac{\vartheta_n}{2}\phi_{n}|v\rangle.
\end{eqnarray}
Here $\Omega_n^R$ and $\vartheta_n$ are defined from
$\Omega_n^R\cos\vartheta_n$=$E_{nc}-E_{nv}$,
$\Omega_n^R\sin\vartheta_n$=$2\sqrt{n}\hbar\omega_{\gamma}$. The
$h_R$ induces the coupling of the LL's, which is schematically shown
in Fig. 2(a). We see that both $\varphi_{nk_x+}^R$ and
$\varphi_{nk_x-}^R$ come from $\phi_{n-1}|c\rangle$ and
$\phi_{n}|v\rangle$. A particular eigenstate is the lowest LL in
valence band ($\phi_0|v\rangle$), which is independent of $h_R$.
Meanwhile, when only $h_D$ exits, we obtain
$E_{n,\pm}^D$=$E_{n,\pm}^R$,
$\psi_{nk_x\pm}^D$=$e^{ik_xx}/\sqrt{L_x}\varphi_{nk_x\pm}^D$
($n$=1,2,...), where
\begin{eqnarray}
\varphi_{nk_x+}^D=\sin\frac{\vartheta_n}{2}\phi_{n}|c\rangle+\cos\frac{\vartheta_n}{2}\phi_{n-1}|v\rangle,\nonumber\\
\varphi_{nk_x-}^D=\cos\frac{\vartheta_n}{2}\phi_{n}|c\rangle-\sin\frac{\vartheta_n}{2}\phi_{n-1}|v\rangle.
\end{eqnarray}
The $h_D$ induces the coupling of the LL's as schematically shown in
Fig. 2(b). We see that both $\varphi_{nk_x+}^D$ and
$\varphi_{nk_x-}^D$ come from $\phi_{n}|c\rangle$ and
$\phi_{n-1}|v\rangle$. A particular eigenstate is the lowest LL in
conduction band ($\phi_0|c\rangle$), which is independent of $h_D$.
Therefore, when both $h_R$ and $h_D$ exist, the LLs are coupled into
the following two groups
\begin{eqnarray}
\phi_0|c\rangle\overset{h_R}{\longleftrightarrow}\phi_1|v\rangle\overset{h_D}{\longleftrightarrow}\phi_2|c\rangle
\overset{h_R}{\longleftrightarrow}\phi_3|v\rangle\overset{h_D}{\longleftrightarrow}\cdots(\text{group}
\; U),
\end{eqnarray}
and
\begin{eqnarray}
\phi_0|v\rangle\overset{h_D}{\longleftrightarrow}\phi_1|c\rangle\overset{h_R}{\longleftrightarrow}\phi_2|v\rangle
\overset{h_D}{\longleftrightarrow}\phi_3|c\rangle\overset{h_R}{\longleftrightarrow}\cdots(\text{group}
\; D),
\end{eqnarray}
The two groups are schematically illustrated in Figs .2(c) and 2(d).
The eigenvalues and eigenvectors can be evaluated numerically by
taking the eigenvectors of $h_c$ in Eq. (7) as basis functions. In
this basis, the wavefunction of the system can be expressed as
\begin{eqnarray}
\psi(x,y)=\frac{e^{ik_xx}}{\sqrt{L_x}}\sum_{m=0}^M\left({\begin{array}{*{20}c} c_m \\
d_m
\end{array}}\right)\phi_m[\kappa(y-y_0)],
\end{eqnarray}
where $\kappa$=$\sqrt{m_{cy}\omega_c/\hbar}$,
$\phi_m(y)$=$e^{-y^2/2}H_m(y)/(\sqrt{\pi}2^m
m!)^{\frac{1}{2}}$ is the harmonic oscillator wave functions.
Then, we can diagonalize the Hamiltonian numerically in a truncated
Hilbert space and obtain the eigenvalues as well as the
eigenvectors.
\begin{figure}
\includegraphics[width=0.5\textwidth]{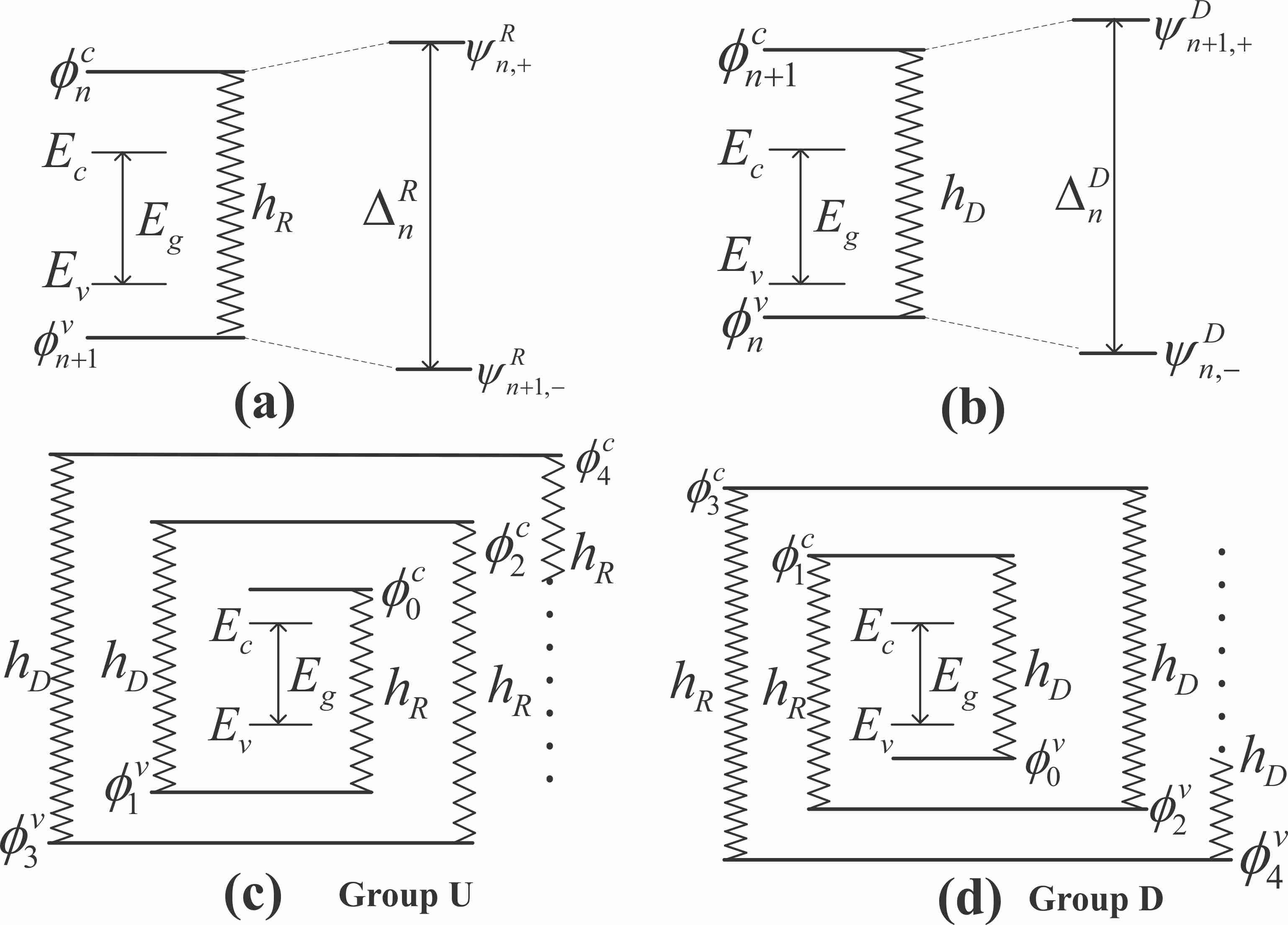}
\caption{Schematic illustration of the inter-LL coupling induced by
(a) $h_R$ with $\Delta_n^R$=$E_{n,+}^R-E_{n+1,-}^R$ and (b) $h_D$
with $\Delta_n^D$=$E_{n+1,+}^D-E_{n,-}^D$; while (c)/(d) represents
coupled LL in group U/D.}
\end{figure}

From the Hamiltonian (2), the LLs can be solved analytically in low
energy regime. Although the dispersion is dominated by the
off-diagonal element, we can decouple the conduction and valence
band in low energy regime due to the large band gap (1.52 eV), i.e.,
the weak interband coupling. The role of the off-diagonal elements
can be taken into account perturbatively. The decoupled Hamiltonian
reads
\begin{eqnarray}
H=\left({\begin{array}{*{20}c}
h_c' &0 \\
0 &h_v'
\end{array}}\right)=\left({\begin{array}{*{20}c}
E_c+\alpha' k_x^2+\beta k_y^2 &0 \\
0 &E_v-\lambda' k_x^2 -\eta k_y^2
\end{array}}\right),
\end{eqnarray}
where $\alpha'$=$\alpha$+$\gamma^2/E_g$,
$\lambda'$=$\lambda$+$\gamma^2/E_g$. The dispersion of this
Hamiltonian is presented by the blue dash-dotted lines in Fig. 1(d).
We see that in the energy regime about 300 meV (see the green solid
line) with respect to the band edges the decoupled Hamiltonian
agrees well with the TB and the \textbf{\emph{k$\cdot$p}} model. The
LL of this Hamiltonian is
\begin{eqnarray}
E_{n,c}=E_c+(n+\frac{1}{2})\hbar\omega_c', E_{n,v}=E_v-(n+\frac{1}{2})\hbar\omega_v',
\end{eqnarray}
where $n$=0,1,2,3,$\cdots$, represents the LL index, the effective
cyclotron frequency
$\omega_c'$=$eB/\sqrt{(m_{cx}'m_{cy})}$=2.657$\omega_e$ and
$\omega_c'$=$eB/\sqrt{(m_{vx}'m_{vy})}$=2.182$\omega_e$ with
$\omega_e$=$eB/m_e$. Note, unlike the anisotropic zero field
dispersion, this LL spectrum is independent on the in-plane
wavevectors. However, the corresponding eignevectors are anisotropic
due to different effective masses along $\Gamma$-$X$ and
$\Gamma$-$Y$ direction. In Landau gauge $\bm{A}$=$(-By,0,0)$, the
eigenvectors are
\begin{eqnarray}
\psi_{n,+}(x,y)=\frac{e^{ik_xx}}{\sqrt{L_x}}\binom{\phi_n(y_c)}0,
\psi_{n,-}(x,y)=\frac{e^{ik_xx}}{\sqrt{L_x}}\binom0{\phi_n(y_v)},
\end{eqnarray}
where $y_{c/v}$=$\kappa_{c/v}(y-y_0)$ with $\kappa_i$=$\sqrt{%
m_{iy}\omega_i^{\prime }/\hbar}$ ($i$=$c,v$). While in
Landau gauge $\bm{A}$=$(0,Bx,0)$, the eigenvectors are
\begin{eqnarray}
\psi_{n,+}(x,y)=\frac{e^{ik_yy}}{\sqrt{L_y}}\binom{\phi_n(x_c)}0,
\psi_{n,-}(x,y)=\frac{e^{ik_yy}}{\sqrt{L_y}}\binom0{\phi_n(x_v)},
\end{eqnarray}
where $x_{c/v}$=$\kappa_{c/v}^{\prime }(x-x_0)$, with $\kappa_i^{\prime }$=$%
\sqrt{m_{ix}^{\prime }\omega_i^{\prime }/\hbar}$ ($i$=$c,v$).
Obviously, the corresponding eigenvectors are anisotropic due to
different effective masses according to Eqs .(15) and (16). Further,
we will see this anisotropy more clearly in symmetry gauge. The wavefunctions in symmetry
gauge are given by
\begin{eqnarray}
\psi_{n,m}(x,y)=A_{n,m}e^{-|Z|^2/2}Z^{|m|}L_n^{|m|}(|Z|^2),
\end{eqnarray}
where $Z$=$X+iY$, $X$=$(x+\delta y)/\sqrt{2\delta}l_B$, $Y$=$(x-\delta y)/\sqrt{2\delta}l_B$, and $\delta$=$\sqrt{m_{cy}/m_{cx}}$ ($\sqrt{m_{vy}/m_{vx}}$) for conduction (valence) band,
and $A_{n,m}$=$(-1)^n\sqrt{n!/(n-m)!\pi}$ is the normalization constant,
$L_n^m(x)$ is the Laguerre polynomials.

In the TB framwork, when the phosphorene sample subjected to a
perpendicular magnetic field, a Peierls phase should be added to the
hopping parameter, which reads
\begin{eqnarray}
H=\sum_{<i,j>}t_{ij}e^{i2\pi\phi_{ij}}c_j^{\dag}c_i,
\end{eqnarray}
where $\phi_{ij}=\frac{e}{h}\int_{r_i}^{r_j}\bm{A}\cdot d\bm{l}$ is
the  Peierls phase. It was first shown by Hofstadter\cite{Hof} that
the energy spectrum in this case depends on a rational dimensionless
parameter $p/q$, where $q$ is a prime number and $p$ runs from 1 to
$q$. This dimensionless parameter is the ratio of magnetic flux
through one unit cell ($\Phi$=$BS$) to the magnetic flux quantum
($\Phi_0$=$h/e$=4.14$\times10^{-15}$T m$^2$), where $S$ is the area
of a unit cell. The energies plotted as a function of $\Phi/\Phi_0$
form a beautiful Hofstadter Butterfly (HB) spectrum. By using Eq.
(17), one will arrive the the Haper's equation and find it is
periodic in $2q$.\cite{Gumbs} The HB spectrum is obtained
numerically by getting the eigenvalues of a matrix with dimension of
$4q\times4q$ at each $\bm{k}$ point in the magnetic Brillouin zone.
A sufficiently large $q$ is needed if one wants to compare this HB
spectrum with the results obtained from the low energy
\textbf{\emph{k$\cdot$p}} model due to the large magnetic flux
quanta.

\begin{figure}
\includegraphics[width=0.5\textwidth]{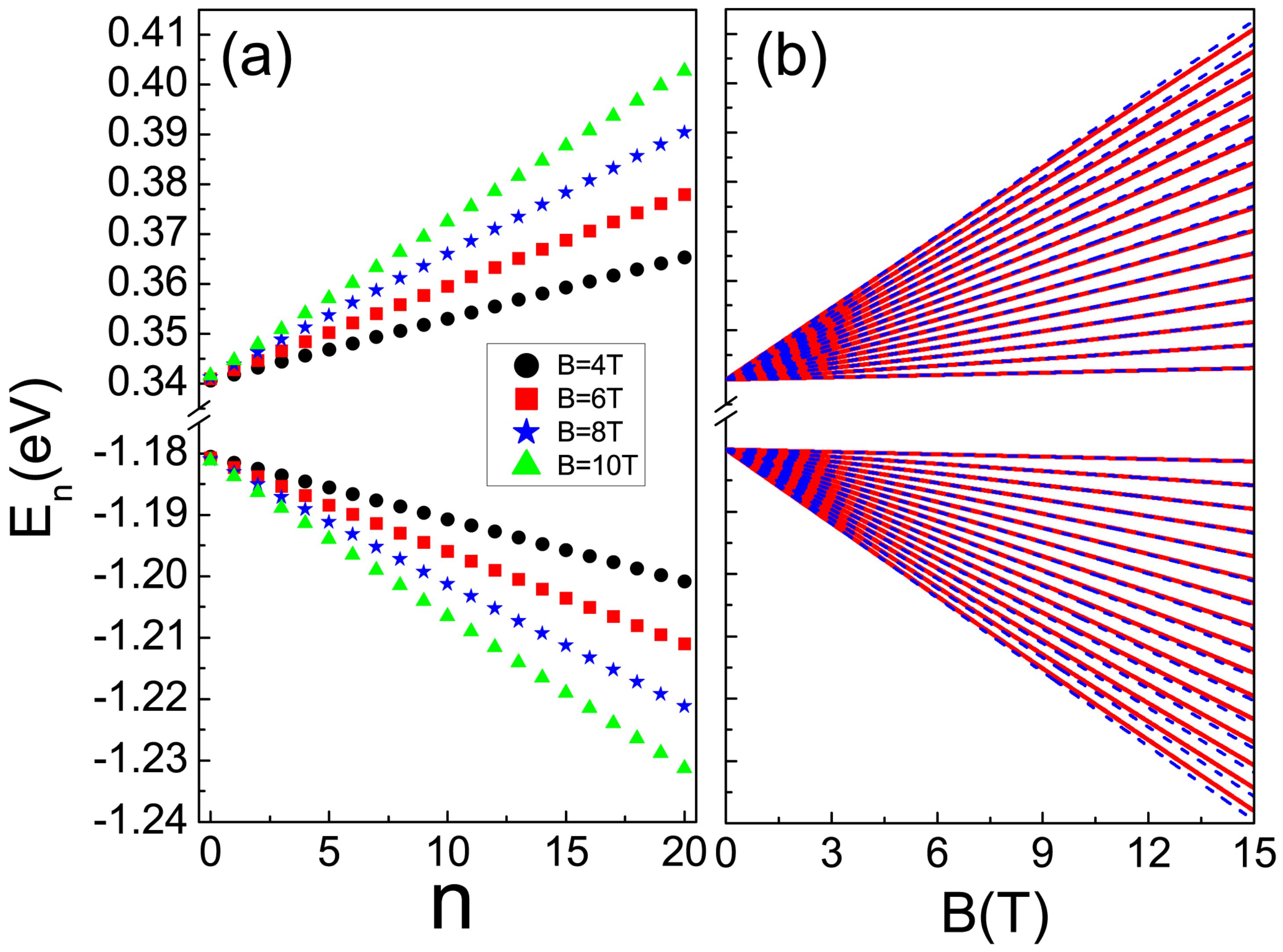}
\caption{(Color online) Landau levels ($E_n$ in units of eV) versus
(a) Landau energy index $n$ with different magnetic field, and (b)
magnetic field $B$ for the first ten low LLs. The number of the
basis function used is 200 to get convergent results. The red solid
lines denote the numerical data and the blue dashed lines represent
the analytical expression in Eq. (14).}
\end{figure}

\subsection{Magneto-transport properties}
In order to detect the calculated magneto energy spectrum, we study
the magneto-transport properties of phosphorene in this subsection.
In the presence of a perpendicular magnetic field, there are two
contributions to magnetoconductance:\cite{Ando} the Hall and
collisional conductance. The former is from the non-diagonal
contribution and the later from the localized states which
contribute to the Shubnikov-de Haas (SdH) oscillation. In order to
calculate the electrical conductance in the presence of a magnetic
field, we follow the formulation of the general Liouville
equation.\cite{Ando} This formulation has been employed successfully
in electron transport for conventional semiconductor
2DEG,\cite{Ando,Yang} and more recently in graphene\cite{Krstajic}
and MoS$_2$.\cite{xyzhou}

Within linear response theory, the Hall conductance in
Kubo-Greenwood formula reads\cite{Ando}
\begin{eqnarray}
\sigma_{\mu \nu}^{\textnormal{nd}}&=&\frac{i\hbar
e^2}{S_0}\sum_{\zetaup\neq\zetaup'}\frac{[f(E_{\zetaup})-f(E_{\zetaup'})]
\langle\zetaup|v_{\mu}|\zetaup'\rangle\langle\zetaup'|v_{\nu}|\zetaup\rangle}{
(E_{\zetaup}-E_{\zetaup'})(E_{\zetaup}-E_{\zetaup'}+i\Gamma_{\zetaup})},
\end{eqnarray}
where $\mu,\nu$$=$$x,y$, $S_0$$=$$L_xL_y$ is the phosphorene sample
area, with the size $L_x$ ($L_y$) in $x$($y$)-direction,
$|\zetaup\rangle$$=$$|s,n,k_x\rangle$ the single electron state in
Eq. (15) as we are interested in the low energy transport,
$f(E_{\zetaup})$$=$$[e^{(E_{\zetaup}-E_F)/k_BT}+1]^{-1}$ the
Fermi-Dirac distribution function with Boltzman constant $k_B$ and
temperature $T$, $v_{\mu}$$=$$\partial H /\partial p_{\mu}$ the
component of group velocity. The sum runs over all states
$|\zetaup\rangle$$=$$|s,n,k_x\rangle$ and
$|\zetaup'\rangle$$=$$|s',n',k_x'\rangle$ with
$\zetaup$$\ne$$\zetaup'$. The infinitesimal quantity
$\Gamma_{\zetaup}$ accounts for the finite broadening of the LLs,
which is assumed approximately the same for all states
\cite{Krstajic}. In our work, we take $\Gamma_{\zetaup}$=0 in order
to obtain a transparent result for Hall conductance. The Hall
conductance is
\begin{eqnarray}
\sigma_{xy}=g_s\frac{e^2}{h}\sum_{n=0,s=\pm}(n+1)[f(E_{n,s})-f(E_{n+1,s})],
\end{eqnarray}
where $g_s$=2 for the spin degree of freedom. At low temperature,
the Hall conductance turns
\begin{eqnarray}
\sigma_{xy}=jg_s\frac{e^2}{h},\;\; (j=0,1,2,3\cdots)
\end{eqnarray}
where $j$ is the filling factor. This result is the same as that for
a conventional semiconductor 2DEG\cite{Ando,Yang} since the zero
field dispersion in low energy regime is quadratic [see Eq. (4)].
\begin{figure}
\includegraphics[width=0.5\textwidth]{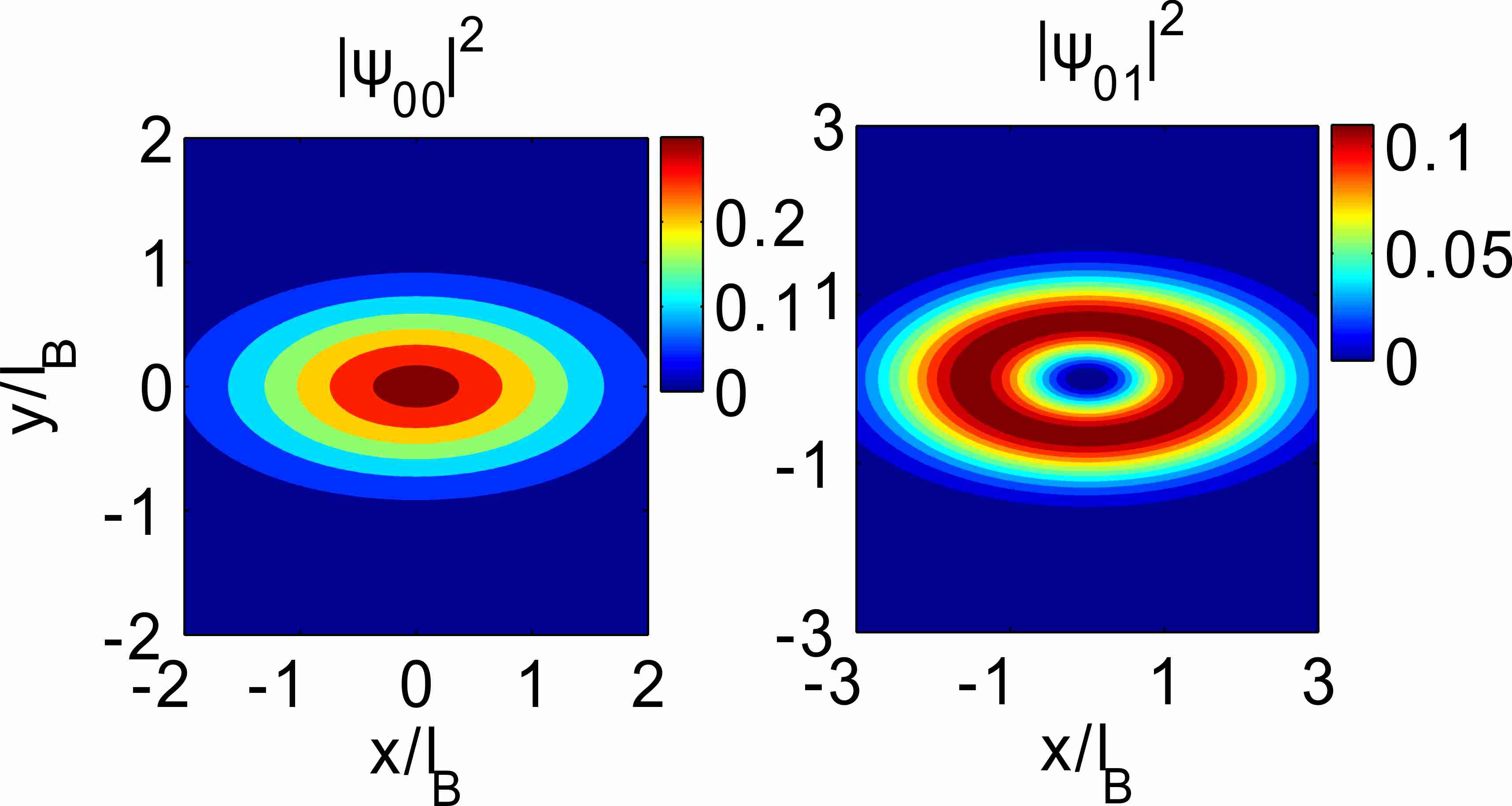}
\caption{(Color online) Contour plot of the spatial density distributions of the
first two LLs in conduction band in symmetry gauge.}
\end{figure}

To obtain the longitudinal conductance, we assume that electrons are
elastically scattered by randomly distributed charged impurities, as
this type of scattering is dominant at low temperatures. The
longitudinal conductance in Kubo-Greenwood formula is given by
\cite{Krstajic}
\begin{eqnarray}
\sigma_{xx}^{\textnormal{col}}=\frac{\beta
e^2}{2S_0}\sum_{\zetaup,\zetaup'}f(E_{\zetaup})[1-f(E_{\zetaup'})]
W_{\zetaup\zetaup'}(E_{\zetaup},E_{\zetaup'})(y_{\zetaup}-y_{\zetaup'})^2,
\end{eqnarray}
where $W_{\zetaup\zetaup'}$ is the scattering rate between
one-electron states $|\zetaup\rangle$ and $|\zetaup'\rangle$.
Conduction occurs via transitions through spatially separated states
from $y_{\zetaup}$ to $y_{\zetaup'}$, where
$y_{\zetaup}$$=$$\langle\zetaup|y|\zetaup\rangle$ is the expectation
value of $y$ coordinate. This means that the longitudinal
conductance arises from the migration of the cyclotron orbit because
of scattering by charged impurities. The scattering rate
$W_{\zetaup\zetaup'}$ is
\begin{eqnarray}
W_{\zetaup\zetaup'}&=&\frac{2\pi n_i}{\hbar
S_0}\sum_q|U_q|^2|F_{\zetaup,\zetaup'}(u)|^2
\delta(E_{\zetaup}-E_{\zetaup'})\delta_{k_x,k_{x}'+q_x},
\end{eqnarray}
where $q$=$\sqrt{q_x^2+q_y^2}$, $u$=$l_B^2q^2/2$, $n_i$ is the
impurity density, $U_q$=$U_0/\sqrt{q^2+k_s^2}$ the Fourier transform
of the screened impurity potential $U(r)$=$U_0e^{-k_sr}/r$ with
$U_0$=$e^2/4\pi \epsilon_0\epsilon_r$, $k_s$ the screening
wavevector, $\epsilon_r$ the dielectric constant and $\epsilon_0$
the dielectric permittivity. Furthermore, if the impurity potential
is strongly short ranged (of the Dirac $\delta$-type function), one
may use the approximation $k_s$$\gg$$ q$ and
$U_q$$\approx$$U_0/k_s$. As the collision is elastic and the
eigenvalue is independent on $k_x$, only the transitions
$n$$\rightarrow$$n$ are allowed. The longitudinal conductance is
\begin{eqnarray}
\sigma_{xx}=g_s\frac{e^2}{h}\frac{n_iU_0^2}{k_BT\hbar\omega_s
k_s^2l_B^2}\sum_{n=0,s=\pm}(2n+1) f(E_{n,s})[1-f(E_{n,s})],
\end{eqnarray}
Moreover, one can obtain the Hall resistance and the longitudinal
one with the conductances ($\sigma_{xy}$ and $\sigma_{xx}$) via
expressions of $\rho_{xy}$$=$$\sigma_{xy}/S$ and
$\rho_{xx}$$=$$\sigma_{xx}/S$, where
$S$$=$$\sigma_{xx}\sigma_{yy}-\sigma_{xy}\sigma_{yx}$
$\approx$$\sigma_{xy}^2$$=$$n_e^2e^2/B^2$,\cite{Ando,Yang,Krstajic}
and $n_e$ is the electron
concentration. For a fixed Fermi energy ($E_f$), $n_e$ is given by
$n_e$=$\int_0^{E_f}D(E)dE$, where $D(E)=\frac{g_s}{\pi l_B^2}\sum_n\delta(E-E_n)$
is the density of states. Note, no matter which
wavefunction [Eq. (15) or (16)] is used in the calculation, one will
obtain the same results as the conductances are gauge independent.

\begin{figure}
\includegraphics[width=0.5\textwidth]{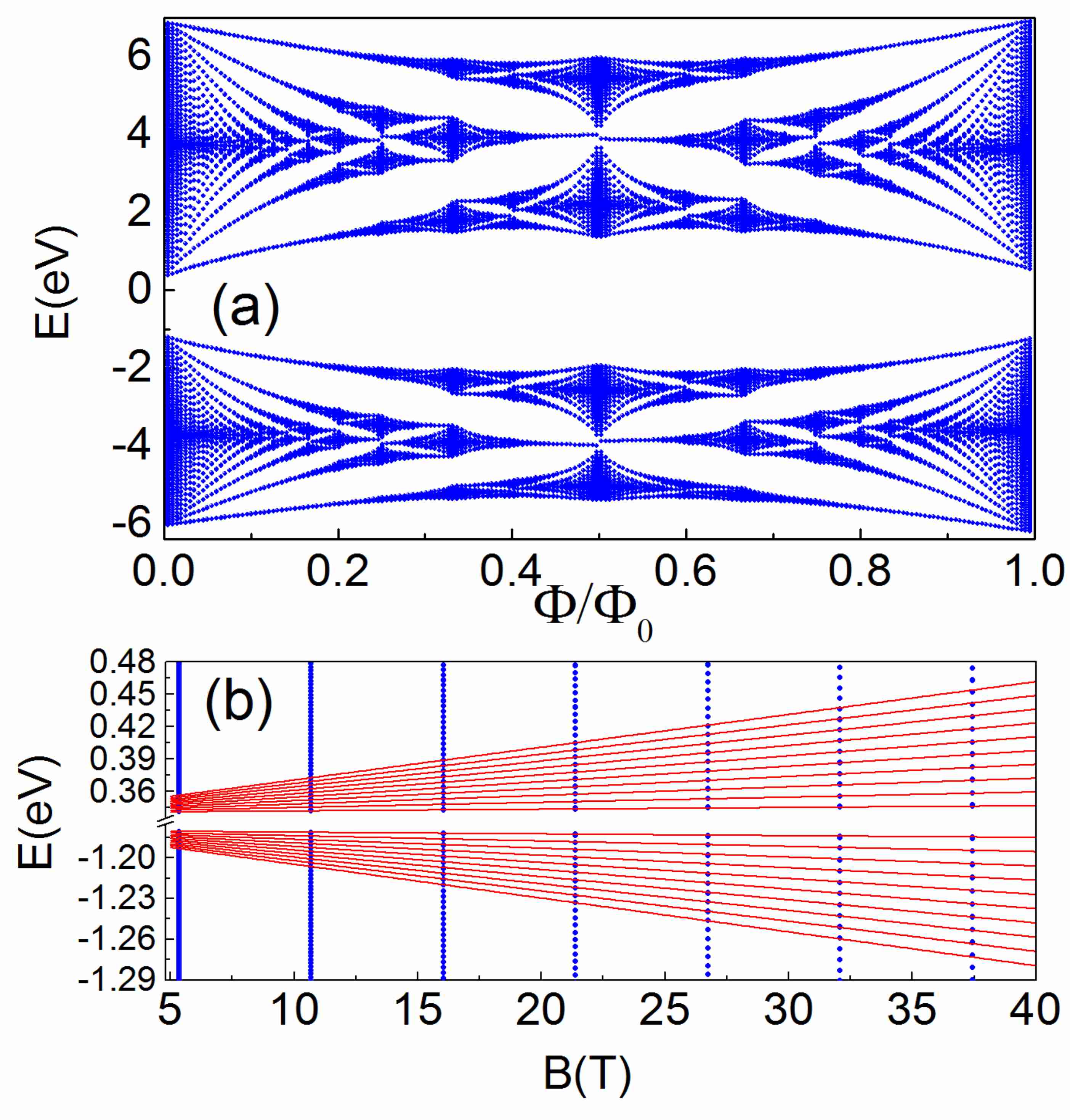}
\caption{(Color online) (a) Hofstadter butterfly (HB) spectrum of phosphorene
with $q$=199 and hopping parameters $t_1$=$-$1.22 eV, $t_2$=3.665
eV, $t_3$=$-$0.205 eV, $t_4$=$-$0.105 eV and $t_5$=$-$0.055 eV. (b) Landau levels obtained from the TB model, i.e.,
the HB spectrum (the blue dots) and the \emph{\textbf{$k\cdot p$}} model (the red solid line) as a function a magnetic
field at low field regime with $q$=10007.}
\end{figure}

\section{Numerical Examples and Discussion}
Next, we show the numerical results with experimental reachable
parameters, i.e., temperature $T$=1 K, impurity concentration
$n_i$=2$\times$10$^{8}$ cm$^{-2}$, screen potential vector
$k_s$=5$\times$10$^{7}$ m$^{-1}$, and dielectric constant\cite{Tos}
$\varepsilon_r$=10.2.

Figure 3 presents the LLs versus (a) LL index $n$ with different
magnetic fields and (b) magnetic field $B$. The red solid lines
denote the numerical data and the blue dashed lines represent the
analytical results in Eq. (14) for the low energy LLs. The number of
basis function used in the calculation is 200 to get convergent
numerical results. As shown in figure 3(b), we find the analytical
LLs (the blue dashed lines) are in good agreement with the numerical
results (the red solid lines), which means the decouple Hamiltonian
(13) is a good approximation in low energy regime. However, the
Landau splittings of conduction and valence band are different for a
fixed magnetic field [see Eq. (14)] due to the different anisotropic
effective masses at zero field. Further, the Landau energies
linearly depend both on LL index $n$ [see Fig. 2(a)] and magnetic
field $B$ [see Fig. 2(b)], which is similar with that of
conventional semiconductor 2DEGs since the zero field dispersion is
quadratic [see Eq. (4)]. Meanwhile, we find the LLs are equally
spaced which can also been seen clearly in Eq. (14).

Arising from the strongly anisotropic band structure at zero field,
the wavefunctions are anisotropic. Figure 4 presents the contour plot of spatial density
distributions (SDDs) corresponding to the first two LLs in conduction
band. As plotted in Fig. 4, unlike the isotropic case,
we find the SDDs for the first two LL are ellipses, which show strong anisotropy.
The decay length of the SDDs
along $x$ direction is larger than that along $y$ direction as the
effective masses along $\Gamma$$-$$X$ direction is smaller than that
in $\Gamma$$-$$Y$ direction [see Eq. (4)] in conduction band. The
same conclusion can be drawn for SDDs corresponding to LLs in valence
band.
\begin{figure}
\includegraphics[width=0.5\textwidth]{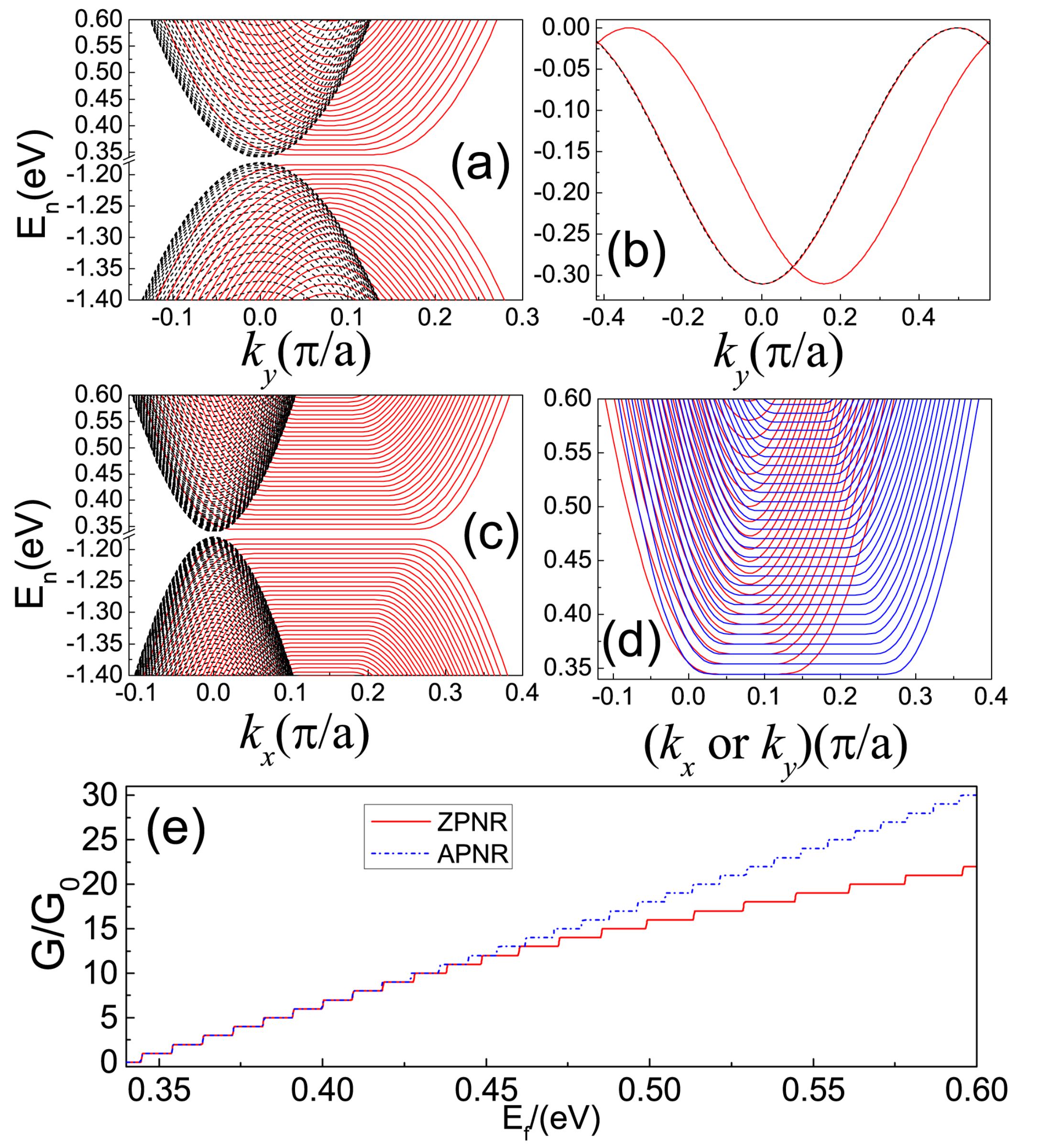}
\caption{(Color online) (a) LLs in a ZPNR with width $w$=66.4 nm, the red solid
 and black dashed lines represent the energy spectra with and without external
 magnetic fields, respectively; $B$=30 T ($l_B$=4.67 nm); (b) The same for (a),
 but for topological flat band located in the bulk gap. (c) LLs in APNR with
 the same parameters used in (a). (d) LLs in ZPNR (red solid lines) and APNR
 (blue solid lines) to show clearly the anisotropic feature of the LLs;
 (e) Conductance (in unit of $G_0$=2$e^2/h$) as a function of Fermi energy for
 ZPNR (red solid line) and APNR (blue-dotted line) corresponding to (d).}
\end{figure}

Adopting the TB model, we plot the Hofstadter Butterfly (HB)
spectrum as a function of $\Phi/\Phi_0$ (magnetic field $B$) with
$q$=199 in Fig. 5(a). As shown in Fig. 5(a), we find two gapped
self-similar HB spectrum coming from the conduction and valence
orbitals, respectively. Moreover, the LL energies linearly depend on
magnetic field $B$ at low field region, which is in line with the
results obtained from the \textbf{\emph{k$\cdot$p}} model [see Eq.
(14)]. The band width of the HB spectrum in conduction and valence
band is different because of the different band widths at zero
field. Figure 5(b) depicts the magneto-levels i.e., the HB spectrum
(the blue dots) and the LLs (the red solid lines) calculated from
the \textbf{\emph{$k$$\cdot$$p$}} theory as a function of magnetic
field at low field regime with $q$=10007. As shown in the figure, we
find they agree well with each other in wide regime of magnetic
fields.

Figure 6 shows the energy spectra of a zigzag-edged PNR
(ZPNR) with and without an external magnetic field. When a strong magnetic
field $B$=30 T is applied perpendicular to the ZPNR, one can clearly see the
LLs. While for an armchair-edged phosphorene nanoribbon (APNR) with the same
with, the LLs show different energy spacing with that in the ZPNR for the
higher LLs. Comparing the energy spectra of the ZPNR and APNR, an important
difference between them is that there is a topological quasi-flat band
located in the bulk gap of the ZPNR\cite{Ezawa}. There are two kinds of
edges states in ZPNRs. The one is the edge states arising from the LLs in
ZPNR, the other come from the topological quasi-flat band. The degeneracy of the topological
qusi-flat band lifts under the influence of the magnetic field. However, since the
topological quasi-flat band are mainly localized near the edges, and the
decay length ($\sim$ 1.2 nm) is less than the magnetic length ($l_B$=25.6nm/$%
\sqrt{B}$=4.67 nm), the edge states arising from the topological quasi-flat
band are almost independent of magnetic fields, i.e., no Landau quantization (see
Fig. 6(b)). The LLs of PNRs depend strongly on the ribbon orientation due to
the anisotropic band structure of bulk phosphorene (see Fig. 6(d)). This
anisotropy of the LLs can be observed in the conductance (see Fig. 6(e)) as
a function of Fermi energy ($E_f$) for ZPNR (red solid line) and APNR (blue
dash-dotted line).

\begin{figure}
\includegraphics[width=0.5\textwidth]{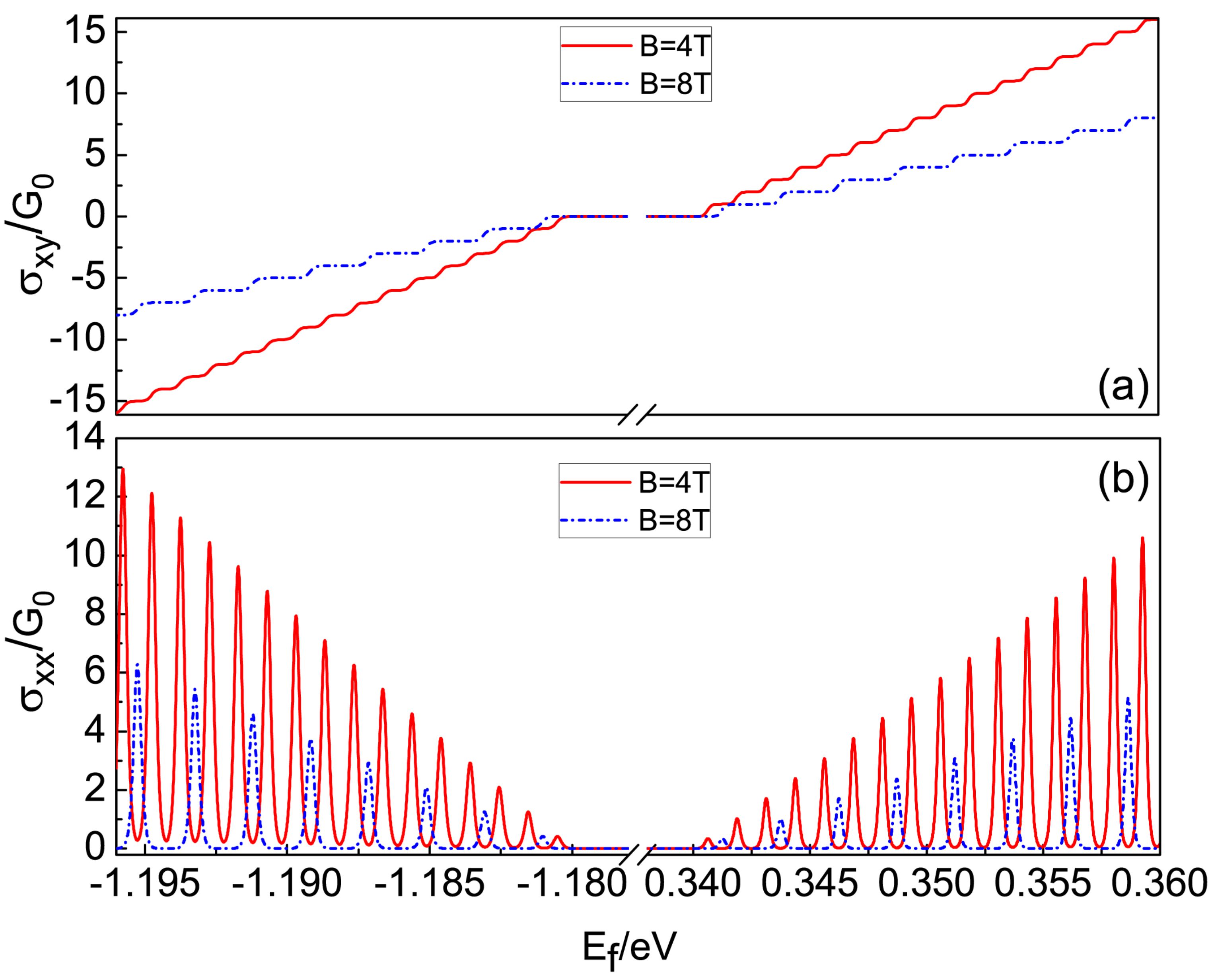}
\caption{(Color online)(a) Hall conductance $\sigma_{xy}$ (in unit
of $G_0=2e^2/h$) and (b) longitudinal conductance versus Fermi
energy $E_f$ (in unit of eV) with different magnetic fields. The
parameters used are: temperature $T$=1 K, impurity concentration
$n_i$=2$\times$10$^{8}$ cm$^{-2}$, screen potential vector
$k_s$=5$\times$10$^{7}$ m$^{-1}$, Boltzman constant
$k_B$=1.38$\times10^{23}$ J/K, and dielectric constant
$\varepsilon_r$=10.2.}
\end{figure}

Figure 7 shows (a) the Hall ($\sigma_{xy}$) and (b) the longitudinal
conductances ($\sigma_{xx}$) as a function of Fermi energy ($E_f$)
for two different magnetic fields $B$=4T and 8T, respectively. As
plotted in Fig. 7(a), we find that Hall conductance is strictly
quantized due to the quantized LLs. It increases one by one in the
unit of $G_0$=2$e^2/h$ with the increasing of Fermi energy since the
LLs are filled one by one. Therefore, we observe the integer Hall
plateaus at 0, $\pm$2, $\pm$4, $\pm$6, $\cdots$ in Hall conductance.
This is similar with that in conventional semiconductor
2DEG.\cite{Yang} Moreover, the Hall conductance reveals the LLs
clearly since the transitions of the plateaus happen to be the
energy value of LLs [see Fig. 3(a)]. Further, for a fixed magnetic
field, the width of the plateaus is equal since the LL spacings of
two adjacent LLs are equal according to Eq. (14). As depicted in
Fig. 7(b), the longitudinal conductance shows that (i) pronounced
peaks appear when the Fermi energy coincides with the LLs, and (ii) a well
splitting SdH oscillation can be observed, which corresponds to the
LLs [see Fig. 3(a)]. Meanwhile, the amplitude of longitudinal
conductance increases with the increasing of the Fermi energy
because of the larger scattering rate of LLs with higher index.
Further, for a given magnetic field, the amplitude of longitudinal
conductance for electrons and holes are slightly different due to
distinct Landau splittings in conduction and valence bands [see Eq.
(14)]. Moreover, the intervals between the peaks are equal since the
LLs are equally spaced according to Eq. (14).
\begin{figure}
\includegraphics[width=0.5\textwidth]{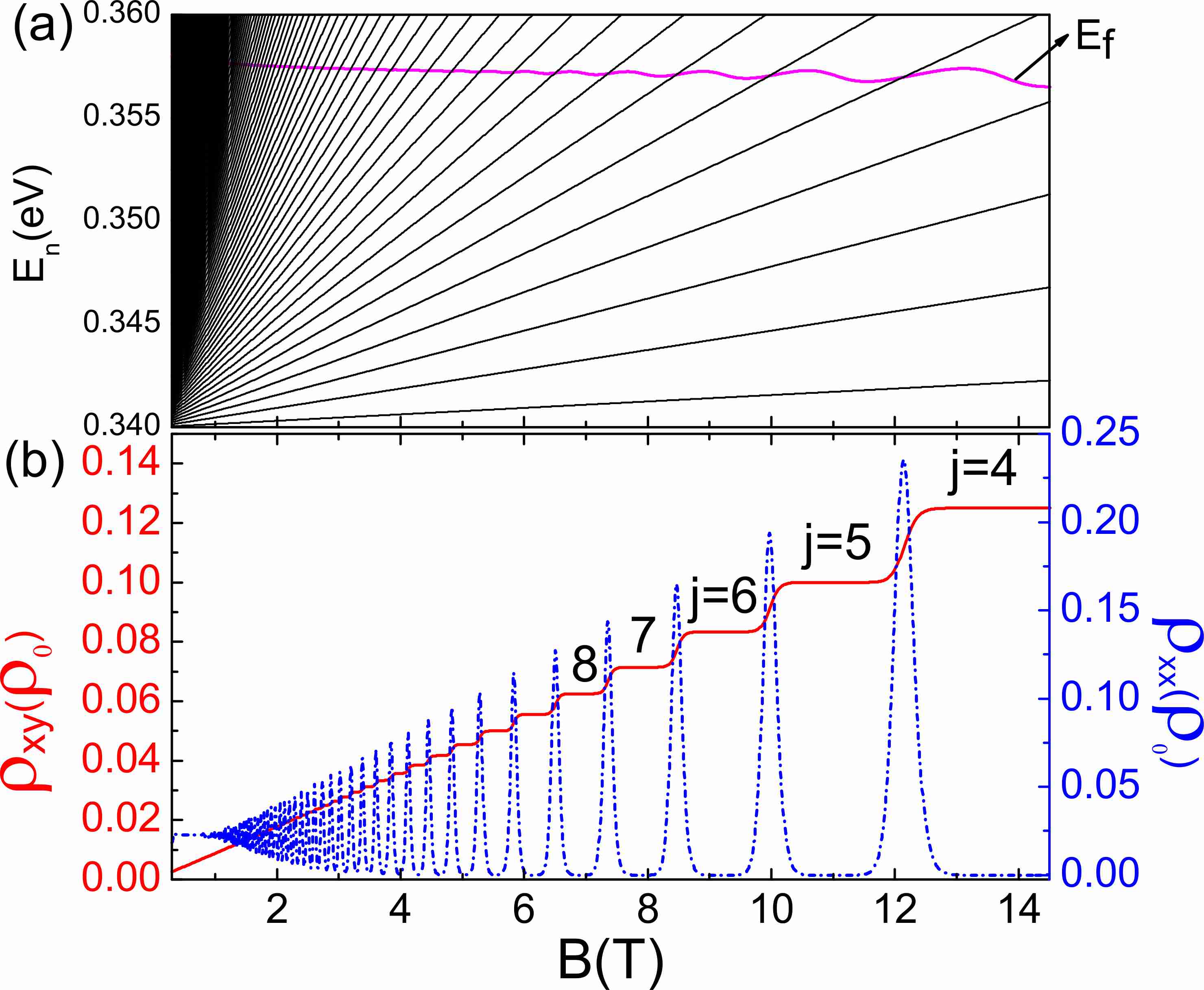}
\caption{(Color online) (a) Electron Fermi energy as a function of
magnetic field for a fixed electron concentration
$n_e$=1.45$\times10^{12} cm^{-2}$, (b) Hall resistance (the red
solid line) and magneto longitudinal resistance (the blue
dash-dotted line) corresponding to (a), the resistance unit $\rho_0$
is $h/e^2$. The other parameters used are the same as Fig. 7. }
\end{figure}

Finally, in Figure 8 we plot the Fermi energy spectra and
resistances as a function of magnetic field for a fixed electron
concentration $n_e$=1.45$\times$10$^{12}$ cm$^{-2}$. Generally, the
Hall ($\rho_{xy}$) and the longitudinal resistances ($\rho_{xx}$)
can be detected directly via Hall measurement.\cite{YBZhang,YBZhange} As
shown in figure 8(b), at low magnetic field, the Hall resistance
linearly depend on the magnetic field and the longitudinal one is a
constant. However, at the high magnetic field regime, the Hall
resistance is strictly quantized with Hall plateaus due to Landau
quantization. It increases (in unit of $\rho_0$=$h/e^2$) one by one
with increasing magnetic field since the LLs leak out of the Fermi
level one by one [see Fig. 8(a)]. This is also reflected in the
transitions of filling factor [see Fig. 8(b)]. Therefore, we observe
plateaus at 1$/$8, 1$/$10, 1$/$12, 1$/$14, $\cdots$, in Hall
resistance corresponding to filling factor $j$=4, 5, 6, 7, $\cdots$,
with the decreasing of magnetic field. Meanwhile, we find a clear
SdH oscillation in longitudinal resistance. The amplitude of
longitudinal resistance increases with the magnetic field since it
is proportional to $B^2$. However, this oscillation is quenched in
low magnetic field due to tiny LL splittings in weak field cases.

\textrm{\\}
\section{Summary}
We studied theoretically the Landau levels and magneto-transport
properties of phosphorene under a perpendicular magnetic field
within the framework of an effective \textbf{\emph{k$\cdot$p}}
Hamiltonian and TB model. In the low field regime, we found that the LLs
linearly depend both on the LL index $n$ and magnetic field $B$,
which is similar with that of conventional semiconductor
two-dimensional electron gas. For a fixed magnetic field, the Landau
splittings of conduction and valence band are different and the
wavefunctions corresponding to the LLs show strong anisotropic
behavior due to the anisotropic effective masses. We obtained an
analytical expression for the LLs in low energy regime via solving a
decoupled Hamiltonian. This analytical solution agrees well with the
numerical results. At high magnetic regime, a self-similar
Hofstadter butterfly (HB) spectrum was obtained by using the TB
model. The HB spectrum is in good agreement with the LLs calculated
from the effective \textbf{\emph{k$\cdot$p}} theory in a wide regime
of magnetic fields.

Further, we found the LLs of phosphorene nanoribbons (PNRs)
depend strongly on the ribbon orientation due to the anisotropic hopping parameters.
There are two kinds of edge states in ZPNRs
under a perpendicular magnetic field. The one is the edge states arising from the LLs,
the other comes from the topological flat band. The second edge states are
almost independent of magnetic fields because their decaying length is less than the magnetic length $l_B$.
Moreover, the Hall  and the longitudinal conductances (resistances) clearly reveal the structure of LLs in phosphorene sheet.

Our conclusions about the LL spectrum in phosphorene can be also applied to multilayer BPs
since the \emph{\textbf{k$\cdot$ p}} Hamiltonians for the multilayer ones are similar with that for phosphorene.
A very recent paper\cite{RXFei} demonstrates that the low energy LLs in bulk phosphorus
also depend linearly both on the LL index $n$ and magnetic field $B$. Meanwhile, this result
has been verified in several recent magneto transport experiments\cite{YBZhange,XLChen,Tayari}.
Our results have been employed to illustrate the absence of non-trivial Berry¡¯s phase of LLs in multilayer BPs\cite{YBZhange}.

\begin{acknowledgments}
This work was supported by the NSFC Grants No. 11434010 and the
grant No. 2011CB922204 from the MOST of China, and NSFC Grants No.
11174252, and No. 11274108.
\end{acknowledgments}

\end{document}